\documentclass[twocolumn,showpacs,preprintnumbers,amsmath,amssymb]{revtex4}
\usepackage{graphicx}
\usepackage{amsmath}
\usepackage{amssymb}
\usepackage{bm}
\usepackage{epstopdf}
\usepackage{slashbox}

\begin{document}
\title{Scaling properties of induced density of chiral and non-chiral Dirac fermions in magnetic fields}
\author{P. S. Park$^{1}$, S. C. Kim$^{1}$, and S. -R. Eric Yang$^{1,2}$\footnote{corresponding author, eyang812@gmail.com}}
\affiliation{
$^{1}$Physics Department, Korea  University, Seoul Korea\\
$^{2}$Korea Institute for Advanced Study, Seoul Korea \\
}
\date{\today}

\begin{abstract}
We find that a repulsive potential of graphene in the presence of a
magnetic field has bound states that are peaked inside the barrier
with tails extending over $\ell(N+1)$, where $\ell$ and $N$ are the
magnetic length and Landau level(LL) index. We have investigated how
these bound states affect scaling properties of the induced density
of filled Landau levels of massless Dirac fermions. For chiral
fermions we find, in strong coupling regime, that the density inside
the repulsive potential can be greater than the value in the absence
of the potential while in the weak coupling regime we find negative
induced density. Similar results hold also for non-chiral fermions.
As  one moves from weak to strong coupling regimes the effective
coupling constant between the potential and electrons becomes more
repulsive, and then it changes sign and becomes attractive.
Different power-laws of induced density are found for chiral and
non-chiral fermions.
\end{abstract}
\maketitle

\section{introduction}

Non-relativistic massless Dirac electrons exist in two-dimensional
graphene layers near K and K${'}$ Brillouin points\cite{Novo,Geim,
Ando, Castro}. Energy dispersions form Dirac cones with conduction
and valence bands meeting at the Dirac point. The wavefunction has
two components: the first component gives the probability amplitude
of finding the electron on A carbon atoms while the second component
gives the amplitude of finding it on B carbon atoms. In the presence
of a magnetic field the Dirac cones split into Landau levels with
some unique features\cite{Gus, Zhang, Miller,Bol}. A Landau level
with zero energy that is independent of magnetic field develops with
chiral wavefunctions. Other non-chiral Landau levels of conduction
and valence bands have opposite energies but their wavefunctions are
identical except for phase factors\cite{Toke, Zheng}. The energies
of these Landau levels depend non-linearly on magnetic field
\begin{eqnarray}
E_N=\textrm{sgn}(N)E_m\sqrt{2|N|}\label{llenergy},
\end{eqnarray}
where the energy separation between these LLs is set by $
E_m=\frac{\hbar v_F}{\ell}$\cite{Sad,Dea}. Valence band LLs are
labeled $N=-1,-2,-3,...$ with decreasing energy, while conduction
band LLs are labeled $N=1,2,3,...$ with increasing energy. The zero
energy LL has $N=0$.

\begin{figure}[!hbpt]
\begin{center}
\includegraphics[width=0.5\textwidth]{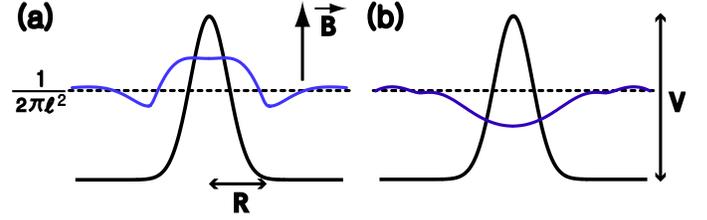}
\caption{ We have plotted schematically electron density for a
filled Landau level. The electron density in the absence of the
potential is $\frac{1}{2\pi\ell^2}$, represented by the dashed
horizontal line. In graphene LLs, under certain conditions, induced
density can be positive (a), in contrast to the usual case where it
is negative (b). }\label{fig:schematic}
\end{center}
\end{figure}

These unusual graphene LLs respond rather differently to presence of
potentials\cite{Chen,Sch,Rec,Park} in comparison with LLs of
ordinary semiconductors : Confinement and deconfinement transitions
\cite{Chen} and bound states forming inside an antidot \cite{Park}
through complete Klein tunneling are predicted. Here we consider the
electron density in the presence of a rotationally invariant and
repulsive potential $V(r)$ with strength $V$ and range $R$. The
electron density of the $N$th filled LL is given by
\begin{eqnarray}
n_N(r)=\sum_{J\in\textrm{filled Landau level}}|\Psi_N^{J}(r)|^2,
\label{density}
\end{eqnarray}
where eigenstates $\Psi_N^{J}(r)$ and eigenvalues $E_N(J)$ are
labeled by LL index $N$ and half-integer angular momentum quantum
number $J$. In the absence of a localized potential the
dimensionless density takes the value $\ell^2n_N(r)=\frac{1}{2\pi}$.
We define the induced density as the difference between densities
with and without the potential
\begin{eqnarray}
\ell^2 \Delta n_N(r)=\ell^2n_N(r)-\frac{1}{2\pi}.
\end{eqnarray}
Graphene barrier has a natural energy scale, $E_c=\frac{\hbar
v_F}{R}$. It is interesting to note that the ratio between this
energy scale and the energy scale of LLs is given by the ratio
between two length scales of the problem:
\begin{eqnarray}
\frac{E_m}{E_c}=\frac{R}{\ell}.
\end{eqnarray}
We find that the correct scaling function of the induced density has
the form
\begin{eqnarray}
\ell^2 \Delta n_N(r)=s_N(\frac{r}{R},\frac{V}{E_c}, \frac{R}{\ell
}).\label{eq2}
\end{eqnarray}
It is different from the scaling function of ordinary LLs where one
would expect that $\frac{V}{E_m}$ appears instead of
$\frac{V}{E_c}$. In graphene $\frac{V}{E_m}$ does not contain
non-perturbative effect of the formation of bound states in the
barrier, and it cannot be used instead of $\frac{V}{E_c}$. Moreover,
the variable $\frac{V}{E_m}$ is inappropriate since it becomes
infinitely  large at zero magnetic field (note $E_m=0$ at $B=0$),
which leads to the unphysical result that the scaling function is
independent of $V$.

\begin{figure}[!hbpt]
\begin{center}
\includegraphics[width=0.52\textwidth]{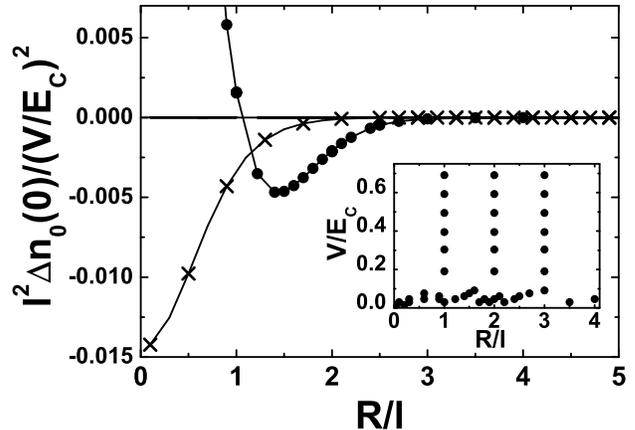}
\caption{ Y-axis represents values of $g_0(R/\ell)$ obtained by data
collapse for N=0 and $\frac{V}{E_{c}}<0.7$. Note that this quantity
is proportional to $-\alpha_N(\frac{V}{E_c},\frac{R}{\ell})$. Values
$(R/\ell,V/E_c)$ used in the data collapse are shown in the inset.
Note that the induced density changes sign near $R/\ell=1$. Crosses
represent the results obtained by treating $V(r)$ in the 2nd order
perturbation theory. These results demonstrate that perturbative
methods cannot be applied in the intermediate and strong coupling
regimes.} \label{fig:scaling0}
\end{center}
\end{figure}

The value of the dimensionless electron density at the center of the
potential ($r=0$) gives a good indication of how strong the {\it
effective coupling constant} between the repulsive potential and
electrons is. We will thus call this dimensionless induced density
at $r=0$, with sign change, as the effective coupling constant
\begin{eqnarray}
 \alpha_N(\frac{V}{E_c},\frac{R}{\ell})=-\ell^2\Delta
n_N(0)=-s_N(0,\frac{V}{E_c}, \frac{R}{\ell }).
\end{eqnarray}
As one moves from weak to strong coupling regimes the effective
coupling constant becomes increasingly more repulsive, and then
starts to decrease, passing through zero, and becomes attractive,
see Figs.\ref{fig:schematic} and \ref{fig:scaling0}. The strong,
intermediate, and weak coupling regimes correspond to
$\frac{R}{\ell}\ll 1$, $\frac{R}{\ell}\sim 1$, and
$\frac{R}{\ell}\gg 1$, respectively. In the strong coupling regime a
repulsive potential can effectively attract electrons, making the
induced density positive. The physical origin of this effect is the
formation of bound states that are peaked inside the barrier with
tails extending over $\ell(N+1)$. We stress that these states are
not resonant states of the repulsive potential in graphene at
B=0\cite{Dong, Sil, Matu} since the extent of the wavefunctions is
finite and their energies form discrete spectra, unlike resonant
states.

In the limit $V/E_c\ll1$ we find the following power laws:
\begin{eqnarray}
\ell^2\Delta n_N(0)\approx(\frac{V}{E_c})^{\delta_N}g_N(R/\ell)
\label{eq2}\label{eq:exx1}
\end{eqnarray}
with  $\delta_N=2$  for the chiral\cite{com1} $N=0$ LL, but with
$\delta_N=1$ for non-chiral $N=1$ LL. We find that $g_0(R/\ell)$
changes sign near 1. A plot of $g_0(R/\ell)$ is shown in
 Fig.\ref{fig:scaling0}.

This paper is organized as follows. In Sec.II we present probability
wavefunctions of boundstates in weak, intermediate, and strong
coupling regimes. Using these probability wavefunctions we
investigate scaling properties of induced densities for chiral and
non-chiral fermions in Sec.III. We also compute the boundary between
positive and negative values of induced densities in the parameter
space. Summary and discussions are given in Sec.IV.

\section{Model calculations of exact wavefunctions }

The electron density of a filled LL is computed from single electron
wave functions, see Eq.(\ref{density}). These wave functions are
chosen to be eigenstates of the following model potential
\begin{eqnarray}
V(r)=\left\{\begin{array}{c}V \ \ \ \ r<R\\
0 \ \ \ \ r>R\end{array}\right. .\label{eqV}
\end{eqnarray}
In the strong coupling regime states with small values of $|J|$ are
peaked near $r=0$ and are particularly relevant for the density
inside of the potential.

\subsection{Energy spectrum }

\begin{figure}[!hbpt]
\begin{center}
\includegraphics[width=0.46\textwidth]{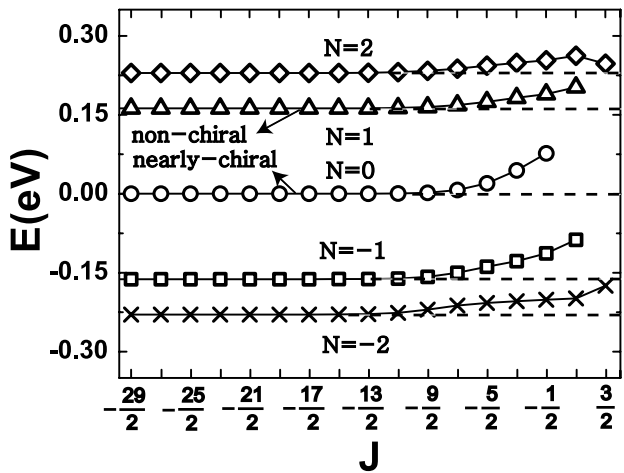}
\caption{ Single particle energy spectrum of massless Dirac fermions
in the presence of repulsive potential and magnetic field. Energies
of five Landau levels are shown for $N=-2,-1,0,1,2$. Dashed lines
represent unperturbed energies, and those that deviate from these
are energies of boundstates. (At $B=20$T, $R=10$nm, $R/{\ell}=1.74$,
$V=0.1$eV).}\label{fig:spectrum}
\end{center}
\end{figure}

\begin{table}[!hbpt]
\caption{Eigenstates are labeled by two quantum numbers: LL index
$N$ and angular momentum quantum number $J$. Possible values of $N$
and $J$ are listed.} \centering
\begin{tabular}{|c|c|}
\hline N & J \\
\hline \vdots & \vdots \\
\hline 2 & $\frac{3}{2}$, $\frac{1}{2}$,$-\frac{1}{2}$,$-\frac{3}{2}$, $-\frac{5}{2}$, $\cdots$\\
\hline 1 &$\quad$$\frac{1}{2}$,$-\frac{1}{2}$,$-\frac{3}{2}$, $-\frac{5}{2}$, $\cdots$\\
\hline 0 &$\qquad$$-\frac{1}{2}$,$-\frac{3}{2}$,$-\frac{5}{2}$, $\cdots$\\
\hline -1 &$\quad$$\frac{1}{2}$,$-\frac{1}{2}$,$-\frac{3}{2}$, $-\frac{5}{2}$, $\cdots$\\
\hline -2 & $\frac{3}{2}$, $\frac{1}{2}$,$-\frac{1}{2}$,$-\frac{3}{2}$, $-\frac{5}{2}$, $\cdots$\\
\hline \vdots & \vdots \\
\hline
\end{tabular}
\end{table}

We compute eigenenergies $E_N(J)$ as a function of $J$ by solving
Dirac equations. We choose the magnetic vector potential as
$\vec{A}=\frac{B}{2}(-y,x,0)$. The two-component wavefunctions of
massless Dirac electrons obey the following Hamiltonian
\begin{eqnarray}
H=v_{F}\vec{\sigma}\cdot(\vec{p}+\frac{e}{c}\vec{A})+V(\vec{r})\label{Eq:Dirac}
\end{eqnarray}
where the Fermi velocity is $v_{F}$, the Pauli spin matrices are
$\vec{\sigma}=(\sigma_{x},\sigma_{y})$. Eigenstates
$\Psi_N^{J}(\vec{r})$ are also eigenstates of angular momentum
operator
\begin{eqnarray}
J_{z}=-i\partial_{\theta}+\frac{\sigma_{z}}{2},
\end{eqnarray}
where $\sigma_{z}$ is a Pauli spin matrix and $\theta$ is the polar
angle. Since $J_z$ commutes with Dirac Hamiltonian of $N$th filled
LL eigenstates must have the following form
\begin{eqnarray}
\Psi^J_{N}(r,\theta)=e^{i(J-1/2)\theta}\left(\begin{array}{c}\chi_{A}(r)\\
\chi_{B}(r)e^{i\theta}\end{array}\right) \label{eq:angular}
\end{eqnarray}
with half-integer values of angular quantum number $J$. For each $N$
allowed values of $J$ are displayed in TABLE I. Some eigenenergies
are shown in Fig.\ref{fig:spectrum}. Although this spectrum
resembles the spectrum of ordinary LLs\cite{Yang1} the wavefunctions
of the eigenstates are rather different.

\subsection{Solutions in strong coupling regime}

\begin{widetext}

\begin{table}[!hbpt]
\caption{How $\psi_i^J$ and $\psi_{n,m}$ are related to each other.
For a given $J$ some possible values of $(n,m)$ are listed. Empty
boxes indicate that $(n,m)$ do not exist.} \centering
\begin{tabular}{|c|c|c|c|c|c|c|c|}
\hline\backslashbox{$\quad$J}{$(n_{i},m_{i})$}  & $(n_{-3},m_{-3})$
& $(n_{-2},m_{-2})$ & $(n_{-1},m_{-1})$
& $(n_{0},m_{0})$ & $(n_{1},m_{1})$ & $(n_{2},m_{2})$ & $(n_{3},m_{3})$ \\
\hline $\frac{3}{2}$ & (-3,1) & (-2,0) & & & & (2,0) & (3,1) \\
\hline $\frac{1}{2}$ & (-3,2) & (-2,1) & (-1,0) & & (1,0) & (2,1) & (3,2) \\
\hline $-\frac{1}{2}$ & (-3,3) & (-2,2) & (-1,1) & (0,0) & (1,1) & (2,2) & (3,3) \\
\hline $-\frac{3}{2}$ & (-3,4) & (-2,3) & (-1,2) & (0,1) & (1,2) & (2,3) & (3,4) \\
\hline
\end{tabular}
\end{table}
\end{widetext}

It is instructive to study properties of probability wavefunctions
in the strong coupling limit, where $R/\ell\rightarrow 0$.
Eigenstates $(\chi_A,\chi_B)$ are determined by the pair of coupled
first order differential equations
\begin{eqnarray}
\left\{\begin{array}{c} -i \partial_x
\chi_A+i\Big[\frac{1}{x}(J-\frac{1}{2})+\frac{1}{2}x\Big]\chi_A=
\epsilon(x)\chi_B\\
-i \partial_x
\chi_B-i\Big[\frac{1}{x}(J+\frac{1}{2})+\frac{1}{2}x\Big]\chi_B=\epsilon(x)\chi_A
\end{array}\right.,
\label{eq1}
\end{eqnarray}
where $x=r/\ell$ is the dimensionless coordinate and
$\epsilon(x)=(E-V(x))/E_m$.

\begin{figure}[!hbpt]
\begin{center}
\includegraphics[width=0.52\textwidth]{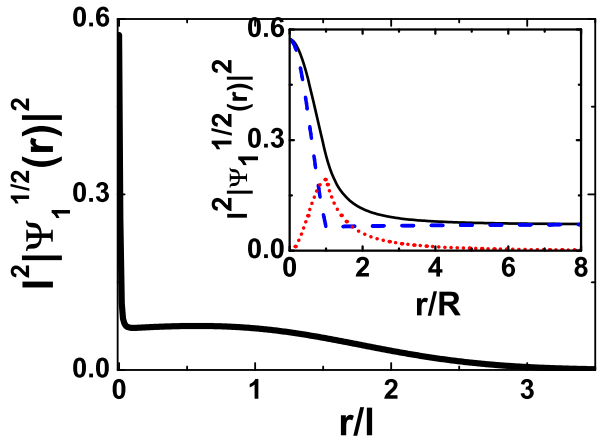}
\caption{ Boundstates of strong coupling limit:
$|\Psi_1^{1/2}(r)|^2$ for $V/E_c=1.82$, $R/\ell=0.01$, $R=1$nm,
$V=1.2$eV, and $E=0.009$eV. Inset: A (B) component is dashed
(dotted) line.}\label{fig:delta}
\end{center}
\end{figure}

For $r\gg R$ the effect of the potential is negligible, and
solutions\cite{Toke} are
\begin{eqnarray}
\psi_i^J(\vec{r})=\psi_{n,m}(\vec{r})=C_{n}\left(\begin{array}{c}-\textrm{sgn(n)}i\phi_{|n|-1,m}(\vec{r})\\
\phi_{|n|,m}(\vec{r})\end{array}\right)\label{twocomp}.\label{eq:spinor}
\end{eqnarray}
Here $n$ and $m$ are integers with $m\geq 0$.
 We define
$\textrm{sgn}(n)=-1,0,1$ for $n<0$, $n=0$ and $n>0$. The
normalization constant $C_n=1$ for $n=0$ and $C_n=1/\sqrt{2}$ for
$n\neq0$. Applying angular momentum operator to Eq.(\ref{eq:spinor})
and comparing with Eq.(\ref{eq:angular}) we find that the quantum
numbers $J$ and $(n,m)$ are related to each other through
\begin{eqnarray}
J=|n|-m-1/2.
\end{eqnarray}
For a given value of $J$ there are infinitely many possible values
of $(n,m)$, and we will order them with an index $i$. How $(J,i)$
are related to $(n,m)$ is given in Table II. In Eq.(\ref{eq:spinor})
the wavefunctions $\phi_{n,m}(\vec{r})$ are the Landau level
wavefunctions of ordinary two-dimensional systems\cite{Yoshi}
\begin{eqnarray}
\phi_{n,m}(\vec{r})&=&A_{n,m}\exp\left(i(n-m)\theta-\frac{r^2}{4\ell^2}\right)\left(\frac{r}{\ell}\right)^{|m-n|}\nonumber\\
&\times&L_{(n+m-|m-n|)/2}^{|m-n|}\left(\frac{r^2}{2\ell^2}\right),
\end{eqnarray}
where $A_{n,m}$ are normalization constants.

To find solutions that are valid for all $r$, we solve the Dirac
equations numerically using confluent hypergeometric
functions\cite{Rec}. The obtained numerical results are shown of
Fig.\ref{fig:delta}. Near $x\geq R/\ell$ the value of the
eigenfunction $\psi_N^J(r)$ is approximately equal to $\psi_i^J(R)$,
given by Eq.(\ref{twocomp}). Both A- and B-components of the
eigenfunction vary rapidly for $r\leq R$. For $J=1/2$  the
A-component of the wavefunction $\chi_A(x)$ is peaked at $x=0$ while
B-component is peaked at $x=R/\ell$. These peak values of  both
$\chi_A$ and $\chi_B$ approach finite values in the limit $R/\ell
\rightarrow 0$. However, as $V/E_c$ increases they also increase.
The amount of jump between $0$ and $R$ is, consistent with
Eq.(\ref{eq1}),
\begin{eqnarray}
-i\Delta
\chi_A=-i(\chi_A(R)-\chi_A(0))\approx\epsilon(0)\int_0^{R/\ell}
 dx
\chi_B
\end{eqnarray}
with $\chi_B(0)=0$ and $\epsilon(0)=(E-V)/E_m$. For $J=-1/2$ the
properties are the opposite to those of $J=1/2$: the B-component of
the wavefunction $\chi_B(x)$ peaked at $x=0$ while A-component is
peaked at $x=R/\ell$.  The amount of jump between $0$ and $R$ is
\begin{eqnarray}
-i\Delta \chi_B\approx\epsilon(0)\int_0^{R/\ell} dx \chi_A.
\end{eqnarray}

\begin{figure}[!hbpt]
\begin{center}
\includegraphics[width=0.45\textwidth]{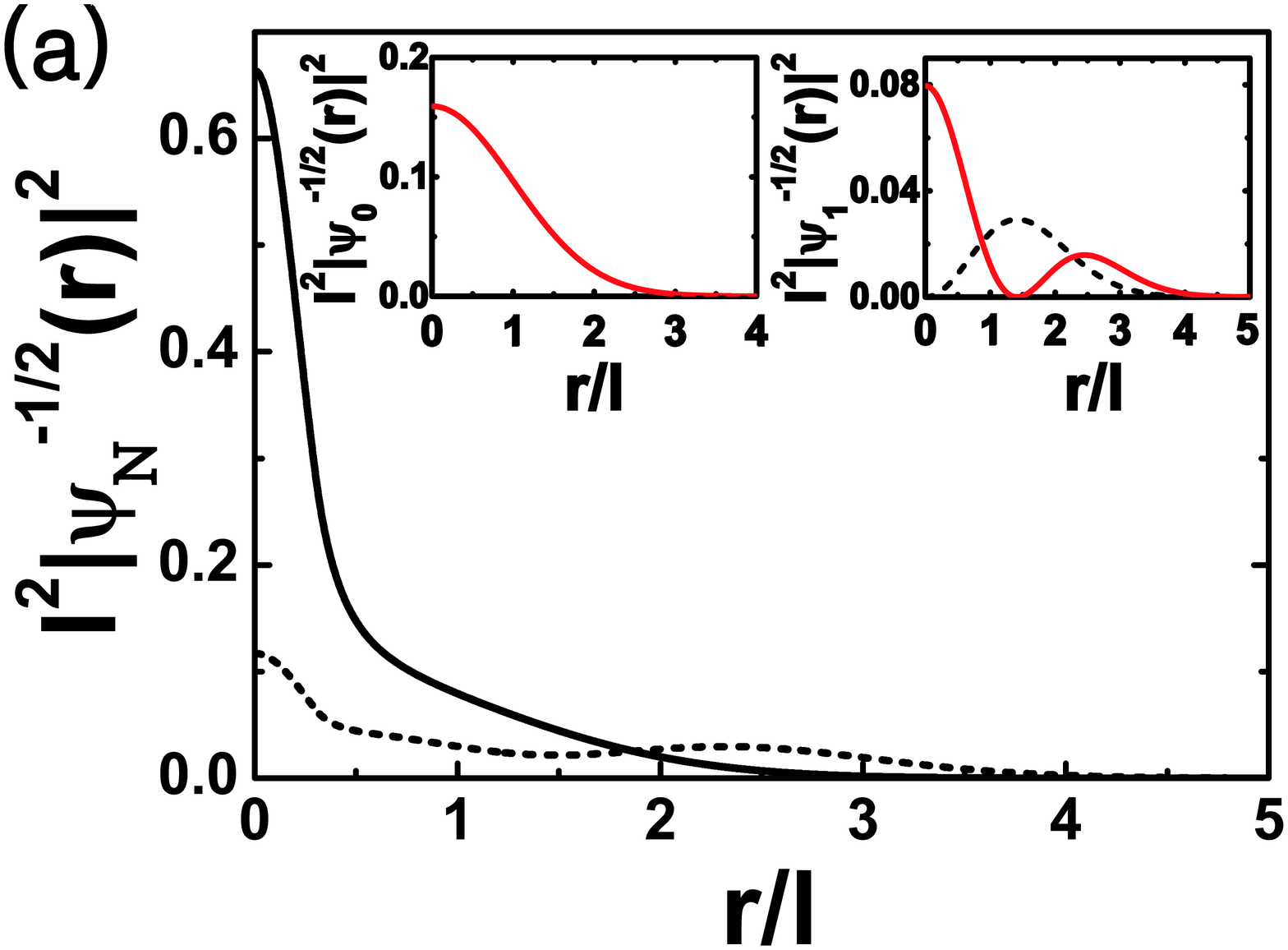}
\includegraphics[width=0.45\textwidth]{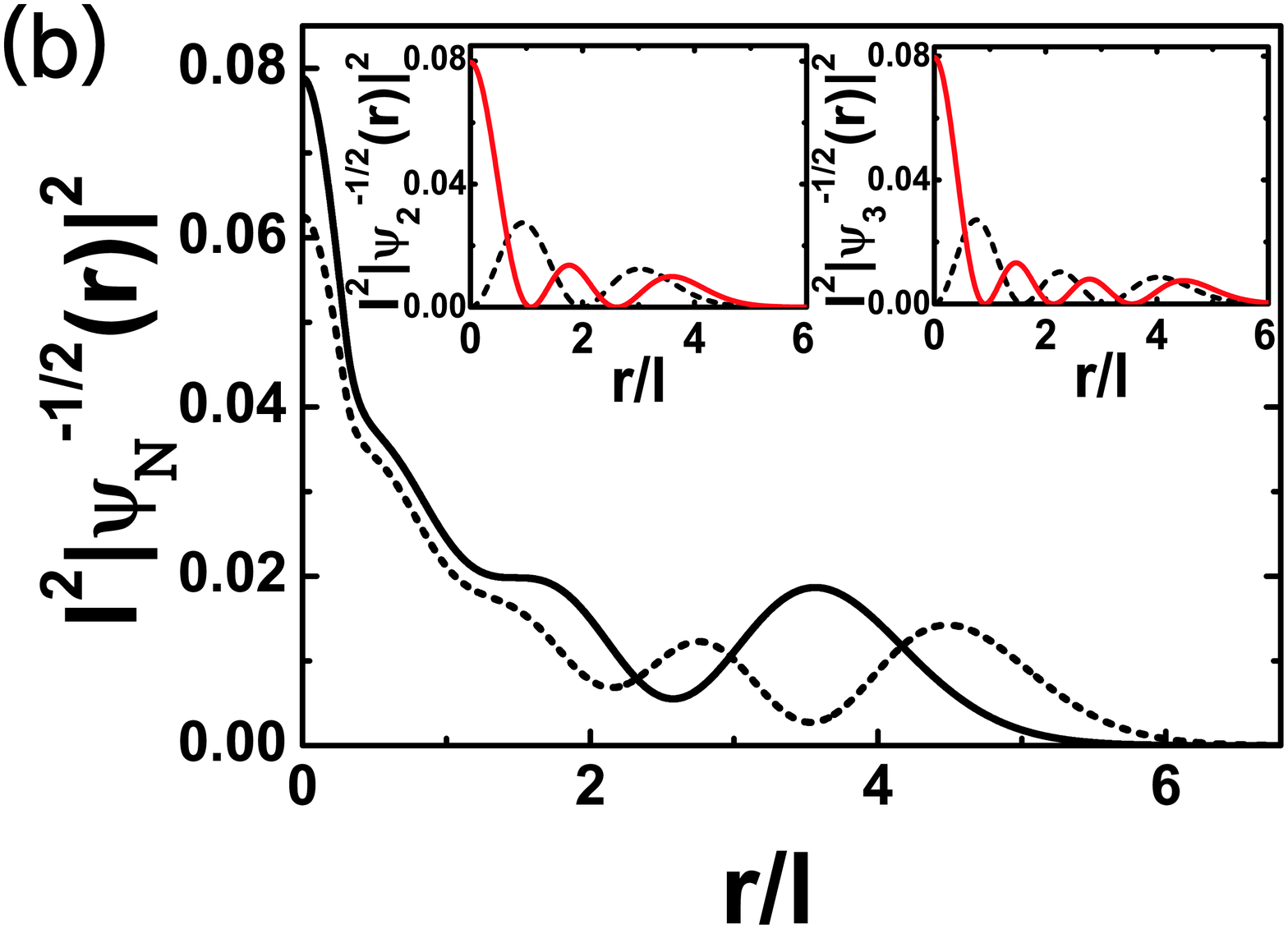}
\caption{ Strong coupling regime: the parameters are $E_c=0.1316$
eV, $E_m=0.0395$ eV, $V=0.26$eV, $V/E_c=1.98$ and $R/\ell=0.3$. (a)
Solid line is for $N=0$ and $E=0.017$ eV.  Dashed line is for  $N=1$
and $E=0.061$ eV. Inset: B-component of $|\Psi_0^{-1/2}(r)|^2$ and
A- (dashed) and B- (solid) components of $|\Psi_1^{-1/2}(r)|^2$ in
the absence of potential. (b) Solid line is for $N$=2 and $E=0.083$
eV. Dashed line is for $N=3$ and $E=0.101$ eV. Inset: A- and
B-components of $|\Psi_2^{-1/2}(r)|^2$ and $|\Psi_3^{-1/2}(r)|^2$ in
the absence of potential. }\label{fig:chiral.vs.nonchiral}
\end{center}
\end{figure}

In the strong coupling regime $R/\ell<1$ probability wavefunctions
can be significant inside the potential, see
Fig.\ref{fig:chiral.vs.nonchiral}. These states are peaked inside
the potential range $R$ and have tails extending over the length of
order $\ell(N+1) $. Examples of such states with $N=0,1,2,$ and $3$
are displayed in Fig.\ref{fig:chiral.vs.nonchiral}. As $N$ increases
the extend of $|\Psi_N^{-1/2}(r)|^2$ outside the potential increases
approximately as $\ell(N+1)$. However, the strength of peak within
the range $R$ decreases with increasing $N$. We stress that these
states are not resonant states of the repulsive potential since the
extend of the wavefunctions is finite and their energies form
discrete spectra, unlike resonant states. States shown in
Fig.\ref{fig:chiral.vs.nonchiral} contribute to a positive induced
charge since the probability of finding an electron inside the
potential has increased compared to the probability in the absence
of the potential. This is a non-trivial effect of the interplay
between effects of quantization of LL and Klein tunneling\cite{Kat,
Sta}.

\begin{figure}[!hbpt]
\begin{center}
\includegraphics[width=0.45\textwidth]{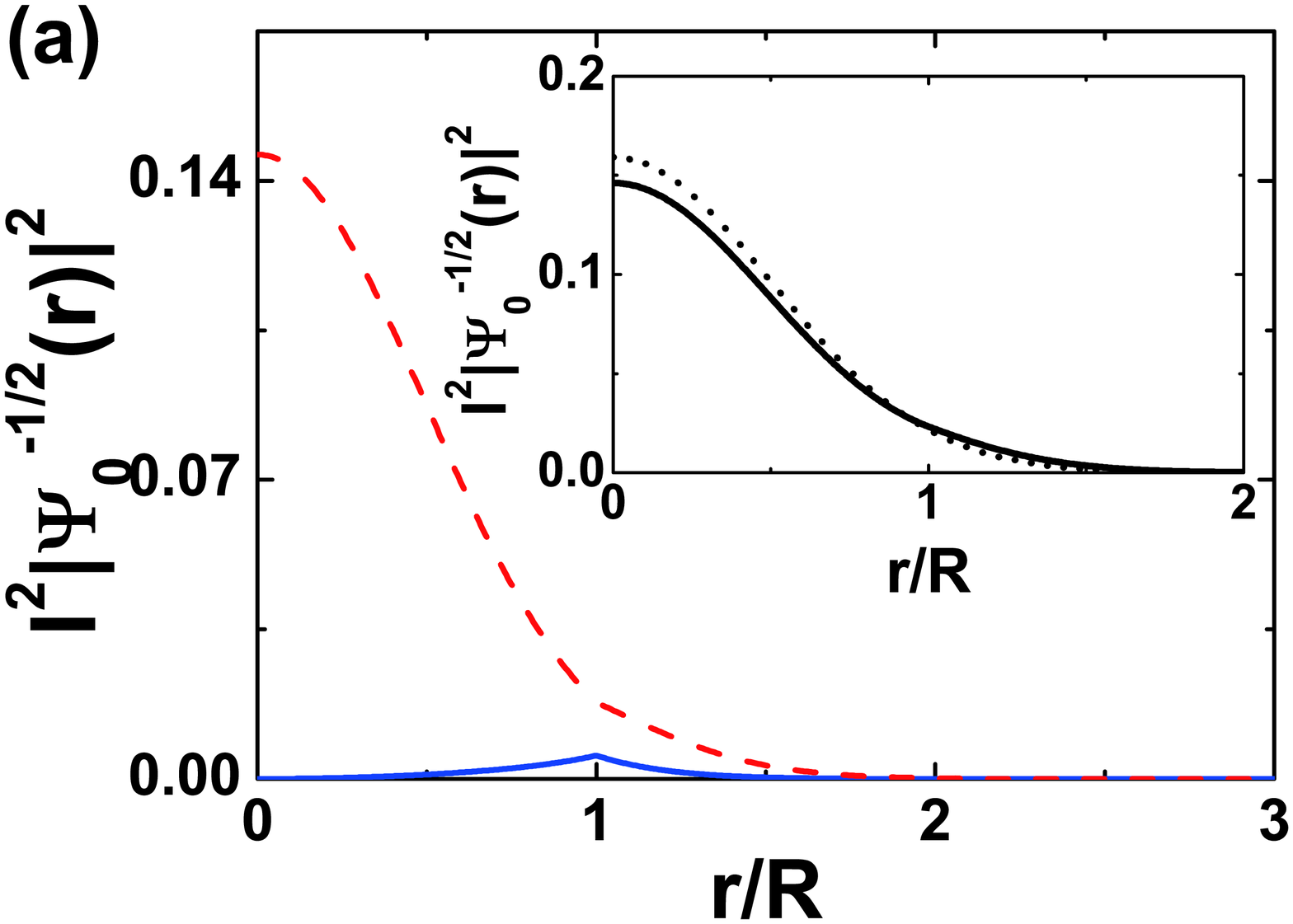}
\includegraphics[width=0.45\textwidth]{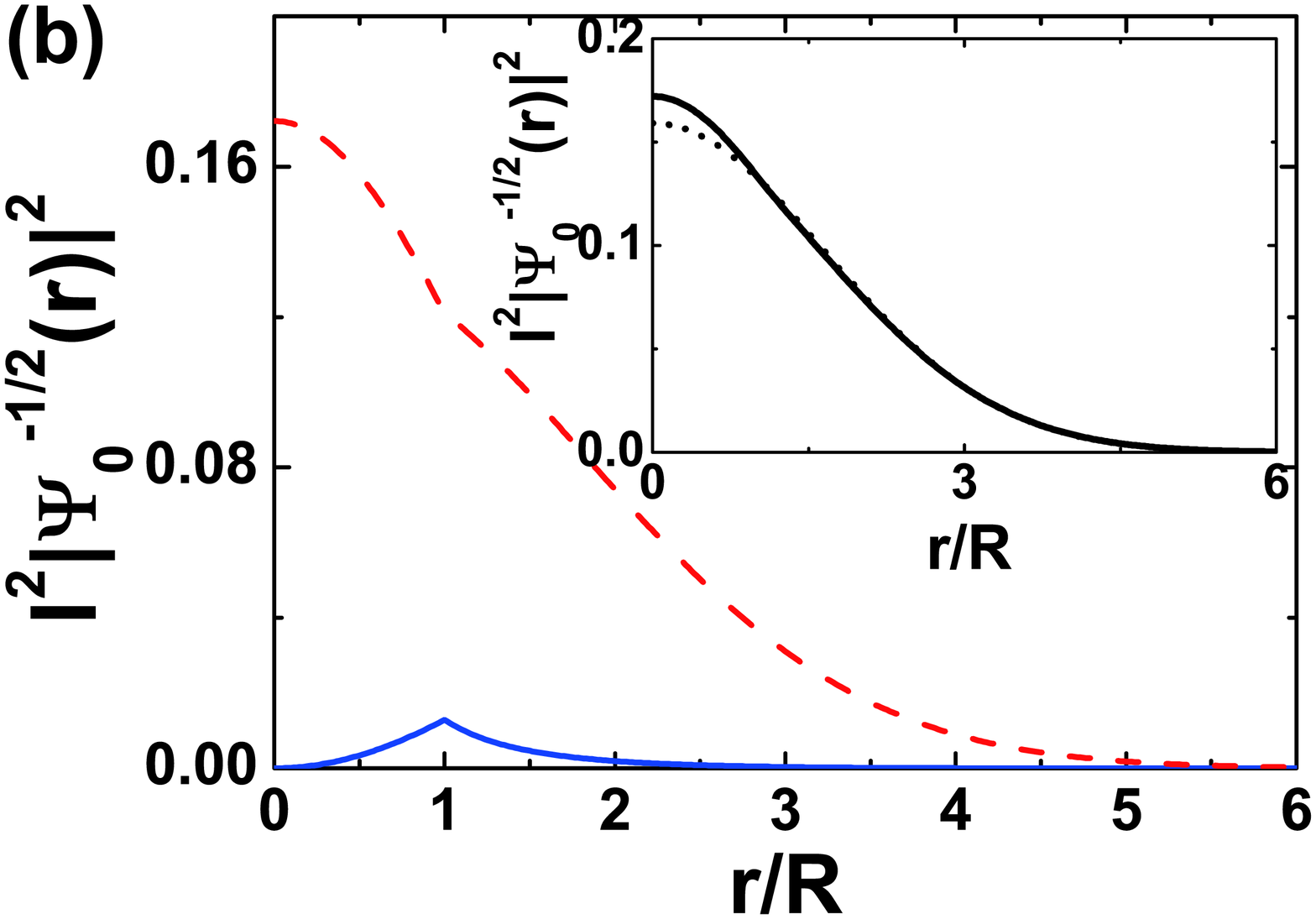}
\caption{ (a) Weak coupling regime: $|\Psi_0^{-1/2}(r)|^2$ for
$R/\ell=2.0$, $V/E_{c}=2.172$, $R=11$nm, $V=0.13$eV, and
$E=0.110$eV. A (B) component is solid (dashed) line. Inset: dashed
line is without potential and solid is with potential. (b)
Intermediate coupling regime: $|\Psi_0^{-1/2}(r)|^2$ for
$R/\ell=0.6$, $V/E_{c}=0.607$, $R=5$nm, $V=0.08$eV, and $E=0.013$eV.
}\label{fig:wavecusp}
\end{center}
\end{figure}

\subsection{Solutions in intermediate and weak coupling regimes}

We show how probability wavefunctions change as $R/\ell$ changes
from weak to intermediate coupling regimes. In the perturbative
regime $R/\ell\gg1$ the exact probability wavefunction at $r=0$ is
smaller than that of the unperturbed probability wavefunction, as
shown in Fig.\ref{fig:wavecusp}(a). Note that cusps in the
wavefunctions are negligible. Fig.\ref{fig:wavecusp}(b) displays the
exact probability wavefunctions  $|\Psi_0^{-1/2}(r)|^2$ in the
intermediate  regime of $R/\ell\sim1$. Both A and B components of it
have cusps at $r=R$. Note that at $r=0$ the exact probability
wavefunction is {\it larger} than that of the unperturbed
probability wavefunction.

\section{scaling function of electron density}

\begin{figure}[!hbpt]
\begin{center}
\includegraphics[width=0.43\textwidth]{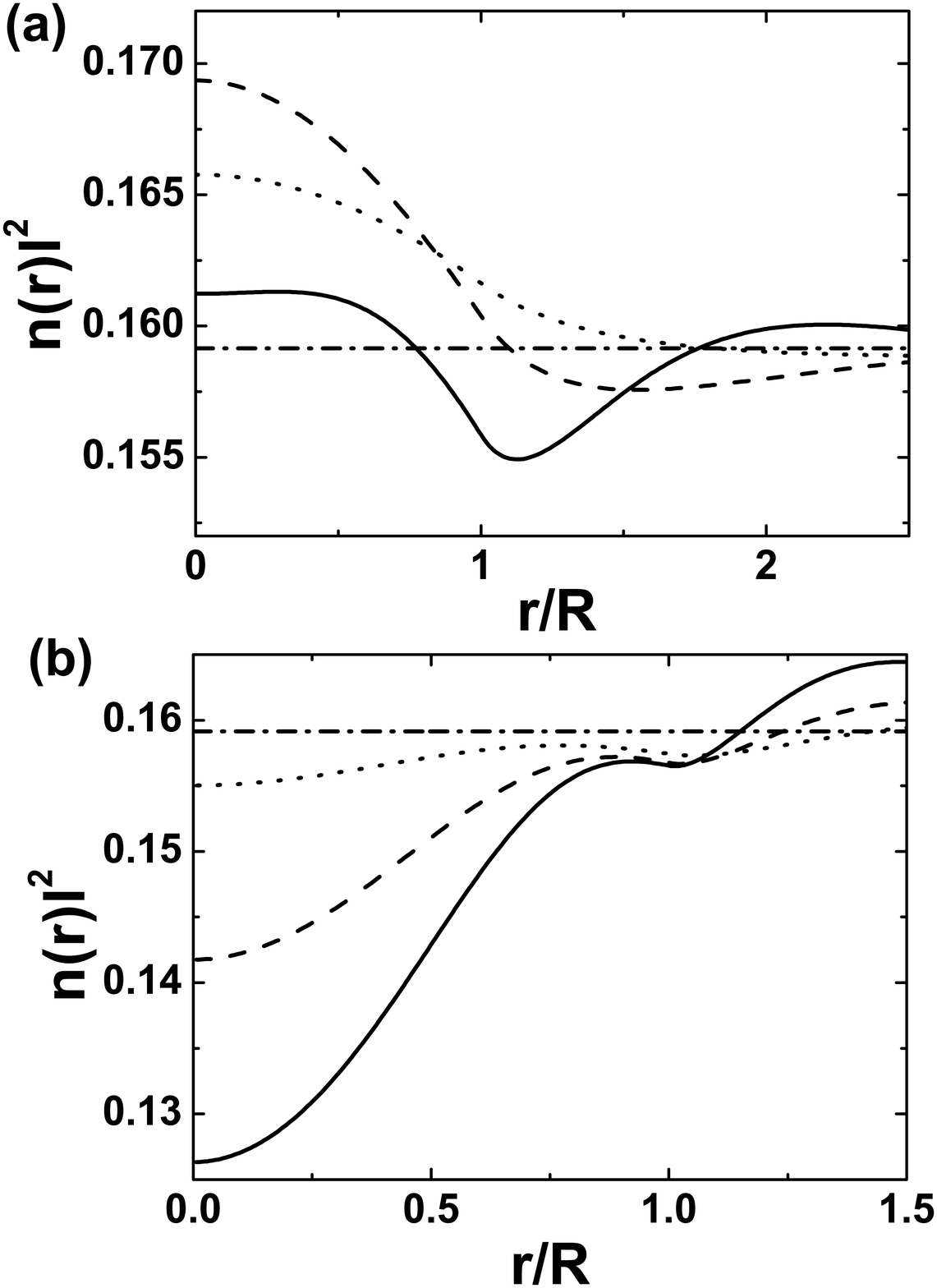}
\caption{(a) $\ell^2 n(r)$ for $N=0$ and  $R/\ell\leq1$.
$(\frac{R}{\ell},\frac{V}{E_{c}})=(1,1.215)$ (solid),
 $(\frac{R}{\ell},\frac{V}{E_{c}})=(0.60, 0.607)$ (dashed),
 $(\frac{R}{\ell},\frac{V}{E_{c}})=(0.30, 0.334)$ (dot).  Dashed-dot line is $\ell^2 n(r)$ in the absence of the potential.
(b) For $N=0$ and  $R/\ell>1$.
$(\frac{R}{\ell},\frac{V}{E_{c}})=(1.74, 2.582)$ (solid),
 $(\frac{R}{\ell},\frac{V}{E_{c}})=(1.50, 1.823)$ (dashed),
 $(\frac{R}{\ell},\frac{V}{E_{c}})=(1.22, 1.063)$ (dot).
}\label{fig:densityn0}
\end{center}
\end{figure}

\begin{figure}[!hbpt]
\begin{center}
\includegraphics[width=0.5\textwidth]{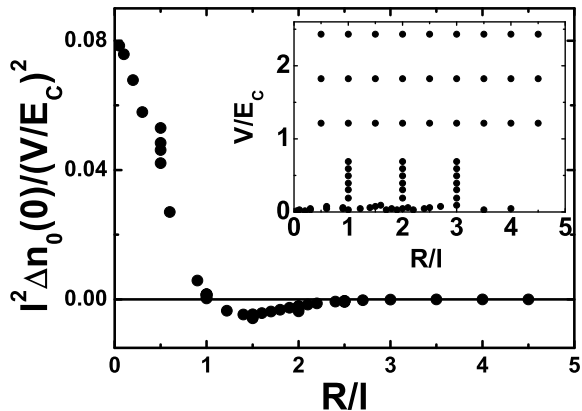}
\caption{ Approximate data collapse of  $g_0(R/\ell)$ is obtained
for a larger range of  $\frac{V}{E_{c}}$ than the one used in
Fig.\ref{fig:scaling0} . Values of $(\frac{V}{E_{c}},R/\ell)$ used
are shown in inset. }\label{fig:scaling00new}
\end{center}
\end{figure}

We have performed an extensive numerical evaluation of the electron
density. The dimensionless induced density for $N=0$ filled LL,
$\ell^2\Delta n_0(r/R)$, is plotted in Fig.\ref{fig:densityn0} for
various values of $(R/\ell,\frac{V}{E_c})$. For $R/\ell<1$ the
induced density inside the barrier is {\it positive}, which is in
sharp contrast to what usually happens in a barrier. The formation
of bound states inside the barrier, as discussed in Sec. II B, is
responsible for this effect. Note the induced density oscillates as
a function of $r/R$. We have tested that the total integrated
density is equal to the total number of electrons in the LL.  The
induced density satisfies the following two-parameter scaling
function
\begin{eqnarray}
\ell^2\Delta n_0(r/R)=s_0(r/R,\frac{V}{E_c},R/\ell). \label{eq:eq2}
\end{eqnarray}

As $R/\ell$ increases the sign of $\Delta n_0(0)$ changes sign from
plus to minus, which is shown in Fig.\ref{fig:densityn0}. Physically
this means positive induced density changes to negative induced
density. For $\frac{V}{E_c}< 0.7$ our numerical results display data
collapse, and is consistent with the following power law
\begin{eqnarray}
\ell^2\Delta n_0(0)\approx(\frac{V}{E_c})^{2}g_0(R/\ell),
\label{eq:exp1}
\end{eqnarray}
see Fig.\ref{fig:scaling0}. For a larger range of
$\frac{V}{E_{c}}< 2.5$
 an approximate data collapse can be obtained, see
Fig.\ref{fig:scaling00new}.

\begin{figure}[!hbpt]
\begin{center}
\includegraphics[width=0.4\textwidth]{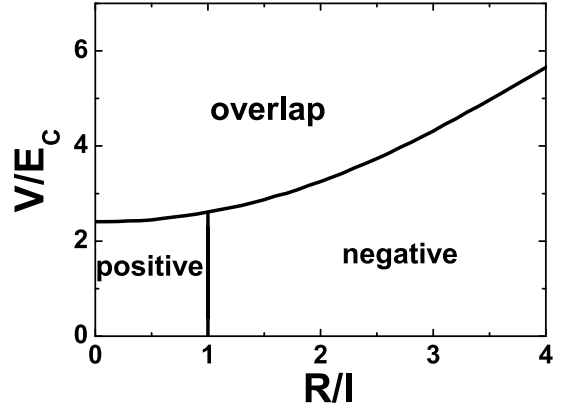}
\caption{ $N=0$ LL states must not overlap with $N=\pm$ 1 LLs.
Induced density at $r=0$ is positive  for
$R/\ell<1$.}\label{fig:overlap2}
\end{center}
\end{figure}

The second order perturbative calculation in $V(r)$ agrees with
scaling result when $R/\ell
>2.5$, but disagrees when $R/\ell <2$, as shown in Fig.\ref{fig:scaling0}.
It is noteworthy that $g_0(R/\ell)$ takes  the minimum value near
$R/\ell\approx 1.5$. This implies that the electron density is
depleted most strongly for $R/\ell\approx 1.5$. However, further
decrease in $R/\ell$ has the opposite effect of increasing more
penetration of electrons into the barrier. Near $R/\ell=1$ the
scaling function $g_0(R/\ell)$ changes sign. For $R/\ell<1$
electrons accumulate in the barrier and the density becomes greater
than the density of the unperturbed LL. This dependence on $R/\ell$
is thus strongly {\it non-linear}. For larger values of $V/E_{c}$
the boundary between positive and negative induced densities is
displayed in Fig.\ref{fig:overlap2}.

\begin{figure}[!hbpt]
\begin{center}
\includegraphics[width=0.42\textwidth]{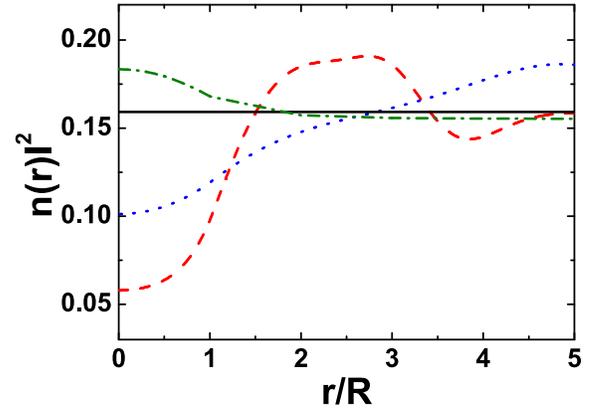}
\caption{$\ell^2 n(r)$ for  $N=1$.
 $(\frac{R}{\ell},\frac{V}{E_{c}})=(1.5, 3.038)$ (dashed),
 $(\frac{R}{\ell},\frac{V}{E_{c}})=(0.6, 0.607)$ (dot),
 $(\frac{R}{\ell},\frac{V}{E_{c}})=(0.01, 0.607)$ (dashed dot).
Solid line is $\ell^2 n(r)$ in the absence of the potential.}
\label{fig:densityn1}
\end{center}
\end{figure}

\begin{figure}[!hbpt]
\begin{center}
\includegraphics[width=0.5\textwidth]{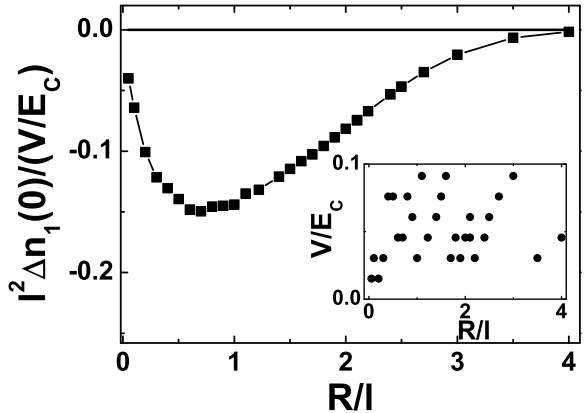}
\caption{
 $g_1(R/\ell)$ is obtained by data collapse  for  $N=1$ and $\frac{V}{E_{c}}\ll1$ .
Values of $(\frac{V}{E_{c}},R/\ell)$ used are shown in inset.
}\label{fig:scaling1}
\end{center}
\end{figure}

\begin{figure}[!hbpt]
\begin{center}
\includegraphics[width=0.42\textwidth]{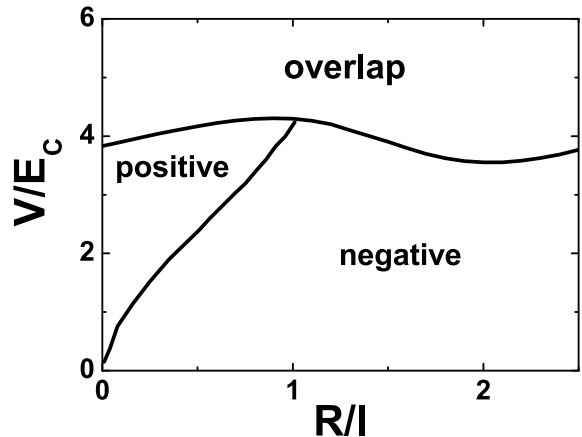}
\caption{ $N=1$ LL states must not overlap with $N=0,2$  LLs. The
range of $R/\ell$, where the induced density at $r=0$ is positive
increases with increasing $V/E_c$.}\label{fig:overlap3}
\end{center}
\end{figure}

 The induced density for $N=1$ filled LL shown in Fig.\ref{fig:densityn1} satisfies a similar scaling relation as that of $N=0$ LL
\begin{eqnarray} \ell^2 \Delta
n_1(r/R)=s_1(r/R,\frac{V}{E_c},R/\ell). \label{eq2}
\end{eqnarray}
At the origin $r=0$ we find, for $\frac{V}{E_c}\ll 1$, the following
scaling result
\begin{eqnarray}
\ell^2\Delta n_1(0)\approx(\frac{V}{E_c})^{\delta_1}g_1(R/\ell),
\label{eq:exp2}
\end{eqnarray}
where $\delta_1=1$. This result is obtained  in the range
$V/E_c<0.1$ by data collapsing numerical data points, see
Fig.\ref{fig:scaling1}. The dependence of the induced density on
$R/\ell$ is again strongly non-linear: $g_{1}(R/\ell)$ takes the
minimum value near $R/\ell=0.7$. For larger values of
$\frac{V}{E_c}$ the boundary between positive and negative induced
densities is displayed in Fig.\ref{fig:overlap3}. As $\frac{V}{E_c}$
increases the range of $R/\ell$ where the induced density is
positive expands. Note that for small values of $R/\ell$, for
example 0.01, the value of $\ell$ at B=1T is $257\AA$ and R becomes
comparable to the lattice constant so that Dirac equations
breakdown. In this case smaller values of B must be used so that R
can take larger values.

It can shown from perturbative analysis that, for chiral fermions,
the first order correction of $V(r)$ absent, but the second order
correction is present and is  negative. The absence of the linear
terms in $\frac{V}{E_c}$ for the chiral $N=0$ LL is a consequence of
the symmetric properties of conduction and valence band LLs.
However, for non-chiral fermions the first order correction is
present. These results are consistent with non-perturbative scaling
results given by Eqs.(\ref{eq:exp1}) and (\ref{eq:exp2}).

We now mention some general properties of the induced density of
LLs. It has a critical point $y_{c,1}$, where
\begin{eqnarray}
\frac{\partial s_N(0,x,y)}{\partial y}\Big|_{y=y_{c,1}}=0
\end{eqnarray}
($x=\frac{V}{E_c}$ and $y=\frac{R}{\ell }$). We find that
$y_{c,1}\sim 1$.  The scaling function takes the global minimum at
$y_{c,1}$: as $y$ decreases the induced density becomes most
negative at $y=y_{c,1}$ and then it increases, in contrast to the
lowest order perturbative result in $V$, which suggests that it
becomes increasingly more negative as $y$ decreases.  Below this
critical point perturbative methods are inapplicable, and it
separates strong and weak coupling regimes. The sign of the induced
density changes at the second critical point $y_{c,2}$, where
\begin{eqnarray}
s_N(0,x,y)|_{y=y_{c,2}}=0.
\end{eqnarray}
As $y$ decreases below $y=y_{c,1}$ the induced density changes sign
at $y=y_{c,2}$ and begins to take positive values.

\section{summary and discussions}

 We find that a repulsive potential of graphene has bound states
that are peaked inside the barrier with tails extending over
$\ell(N+1)$. The properties of these boundstates change as $R/\ell$
varies, and  affect  the induced density of filled LLs inside the
barrier in a non-trivial way as a function of $R/\ell$. As $R/\ell$
decreases the induced density inside the barrier becomes more
negative, but as it reaches a critical value the induced density
reaches an extremum value. Upon further decrease of $R/\ell$ the
value of the induced density reaches zero, and, after this, it
becomes positive. These changes are strongly {\it non-linear} in
$R/\ell$, and one moves successively from weak, intermediate, and
strong coupling regimes as $R/\ell$ decreases. The condition
$V/E_c\ll 1$ is not sufficient for treating $V(r)$ perturbatively,
and, in addition to this, one must require $R/\ell\gg 1$  for both
chiral and non-chiral fermions.

For filled LLs electron-electron interactions may be approximated
well by a Hartree-Fock (HF) method\cite{Mac}. The electron density
in HF method  is given by the sum of unrenormalized single electron
probability wavefunctions, just as in Eq.(\ref{density}). So our
calculation of the induced density is actually a HF result. However,
our single electron energies are not the renormalized HF result.
This will somewhat affect our numerical estimate of the boundary
between overlapping LLs of Figs.\ref{fig:overlap2} and
\ref{fig:overlap3}. The discontinuity in the  potential of
Eq.(\ref{eqV}) can couple states in K and K${'}$ valleys, which is
ignored in our approach. However, our tight-binding calculations
show that this coupling is small\cite{Park}.

There is a symmetry\cite{Park} between repulsive and attractive
potentials $V(r)$ and $-V(r)$ so that  the induced densities of
these potentials are identical. In the presence of a repulsive or
attractive  potential, both charge accumulation and depletion occur,
depending on the value of $R/\ell$, see Figs.\ref{fig:overlap2} and
\ref{fig:overlap3}. The appearance of a charge accumulation near a
repulsive potential, for example, could be explained by introducing
an attractive potential via the transformation $V(r)\rightarrow
-V(r)$, but this same transformation would fail to explain charge
depletion since electrons would pile up around the transformed
attractive potential. Thus charge depletion and accumulation cannot
be explained simultaneously in either perspective of repulsive or
attractive potential. In addition, the critical points
$R/\ell=y_{c,1}$ and $y_{c,2}$, where the scaling function takes the
global minimum and where it changes sign cannot be explained by
application $V(r)\rightarrow -V(r)$. It would be desirable to
construct an analytic theory for them.

Properties of the bound states of the potential barrier may be
observed as follows. A localized potential may be created by a
circular gate placed on graphene sheet.  When this gate is
sufficiently close to the edge of the sample coupling between bound
states  and edge states may be induced, and the transmission
coefficients of edge states may reveal properties of the bound
states.  Also these boundstates may play an important role in
transport and magnetic properties of graphene\cite{Huang}.

S.R.E.Y. thanks Philip Kim for valuable discussions on various
aspects of this paper. This work was supported by the Korea Research
Foundation Grant funded by the Korean Government (KRF-2009-0074470).


\begin{references}
\bibitem{Novo} K. S. Novoselov, A. K. Geim, S. V. Morozov, D.
Jiang, Y. Zhang, S. V. Dubonos, I. V. Grigorieva, A. A. Firsov,
Science, {\bf 306}, 666 (2004).
\bibitem{Geim} A. K. Geim and A. H. MacDonald, Phys. Today, {\bf 60}(8), 35 (2007).
\bibitem{Ando} T. Ando, J. Phys. Soc. Jpn. {\bf 74}, 777
(2005).
\bibitem{Castro} A. H. Castro Neto, F. Guinea, N. M. R. Peres, K. S.
Novoselov, and A. K. Geim, Rev. Mod. Phys. {\bf 81}, 109 (2009).
\bibitem{Gus} V. P. Gusynin, and S. G. Sharapov, Phys. Rev. Lett. {\bf 95}, 146801 (2005).
\bibitem{Zhang} Y. Zhang, Y. W. Tan, H. L. Stormer, P. Kim, Nature, {\bf
438}, 201 (2005).
\bibitem{Miller} D. L. Miller, K. D. Kubista, G. M. Rutter, M. Ruan, W. A. de Heer, P. N. First, and J. A. Stroscio, Science, {\bf 324}, 924 (2009).
\bibitem{Bol} K. I. Bolotin, F. Ghahari, M. D. Shulman, H. L. Stormer, and P. Kim, Nature, {\bf
462}, 196 (2009).
\bibitem{Zheng} Y. Zheng and T. Ando, Phys. Rev. B {\bf 65}, 245420
(2002).
\bibitem{Toke} C. T\H{o}ke, P. E. Lammert, V. H. Crespi, and J. K. Jain, Phys. Rev. B {\bf 74}, 235417
(2006).
\bibitem{Sad} M. L. Sadowski, G. Martinez, M. Potemski, C. Berger, and
W. A. de Heer, Phys. Rev. Lett. {\bf 97}, 266405 (2006).
\bibitem{Dea} R. S. Deacon, K. -C. Chuang, R. J. Nicholas, K. S.
Novoselov, and A. K. Geim, Phys. Rev. B {\bf 76}, 081406R (2007).
\bibitem{Chen} H. Y. Chen, V. Apalkov, and T. Chakraborty, Phys. Rev. Lett.
{\bf 98}, 186803 (2007); G. Giavaras, P.A. Maksim, and M. Roy, J.
Phys.: Condens. Matter {\bf 21}, 102201 (2009).
\bibitem{Sch} S. Schnez, K. Ensslin, M. Sigrist, and T. Ihn, Phys. Rev. B {\bf
78}, 195427 (2008).
\bibitem{Rec} P. Recher, J. Nilsson, G. Burkard, B. Trauzettel, Phys. Rev. B {\bf
79}, 085407 (2009).
\bibitem{Park} P. S. Park, S. C. Kim, and S.-R. Eric Yang, J. Phys.:
Condens. Matter {\bf 22}, 375302 (2010).
\bibitem{Dong}S.-H. Dong, X.-W. Hou, and Z.-Q. Ma,  Phys. Rev. A {\bf 58}, 2160, (1998).
\bibitem{Sil} P. G. Silvestrov and K. B.
Efetov,  Phys. Rev. Lett. {\bf 98}, 016802, (2007).
\bibitem{Matu} A. Matulis and F. M. Peeters, Phys. Rev. B {\bf 77},
115423 (2008).
\bibitem{com1}It should be noted that a localized potential couples chiral LL states
to non-chiral LL states, and thus makes  the $N=0$ band states
slightly non-chiral.  Since these states are dominantly chiral we
will call them chiral states. In $N=1$ states  both A and B
components are present, and these states are  non-chiral.
\bibitem{Yang1} S. -R. Eric Yang and A. H. MacDonald, Phys. Rev. B, {\bf 42}, 10811R (1990).
\bibitem{Yoshi} D. Yoshioka, {\it The Quantum Hall Effect} (Springer, Berlin,
1998).

\bibitem{Kat} M. I. Katsnelson, K. S. Novoselov, and A. K. Geim, Nature Phys. {\bf 2}, 620 (2006).
\bibitem{Sta} N. Stander, B. Huard, and D. Goldhaber-Gordon,  Phys. Rev. Lett. {\bf 102},
026807 (2009).
\bibitem{Mac} A. H. MacDonald, S. -R. Eric Yang,
and M. D. Johnson, Aust. J. Phys. 46, 345
 (1993); S. -R. Eric Yang, A. H. MacDonald, and M. D. Johnson, Phys. Rev. Lett. {\bf 71}, 3194 (1993).
\bibitem{Huang} B.-L. Huang, M.-C. Chang, and C.-Y. Mou, Phys. Rev. B, {\bf 82}, 155462 (2010);
H. Ohldag, T. Tyliszczak, R. Hohne, D. Spemann, P. Esquinazi, M.
Ungureanu, and T. Butz,  Phys. Rev. Lett. {\bf 98}, 187204 (2007).
\end{references}
\end{document}